\documentstyle[epsf]{article}
\include{epsf}
\begin{document}
\centerline{\Large \bf Effect of silica colloids on the rheology
of} \vspace{0.25cm} \centerline{\Large \bf viscoelastic gels
formed by the surfactant} \vspace{0.25cm} \centerline{\Large\bf
cetyl trimethylammonium tosylate} \vspace{0.5cm} \centerline{\bf
Ranjini Bandyopadhyay \footnote{presently at the Department of
Physics and Astronomy, Johns Hopkins University, Baltimore MD
21218, USA, Email: ranjini@pha.jhu.edu} and A. K. Sood}
\vspace{0.75cm} \centerline{\it  Department of Physics, Indian
Institute of Science, Bangalore 560 012, India }
\baselineskip=24pt
\vspace{1.5cm}

\centerline{\Large \bf Abstract} \noindent The effects of the
addition of sub-micrometer sized colloidal silica spheres on the
linear and nonlinear rheology of semi-dilute solutions of a
viscoelastic gel are studied. For a 1.4 wt.\% solution of the
surfactant CTAT, a peak in the zero shear rate viscosity
$\eta_{\circ}$ is observed at approximately equal weight percents
of silica and CTAT. This peak shifts to lower silica
concentrations on increasing either the CTAT concentration or the
surface charge on silica and disappears when the CTAT
concentration is increased to 2.6wt\%. The increases in
$\eta_{\circ}$ and the high frequency plateau modulus G$_{\circ}$
 on the introduction of SiO$_{2}$ are explained by considering
 the increasingly
entangled wormlike micelles that are formed due to the enhanced
screening of the electrostatic interactions. The observed decrease
in the values of G$_{\circ}$ and $\eta_{\circ}$ at higher
concentrations of silica particles is explained in terms of the
formation of surfactant bilayers due to the adsorption of the
positively charged cetyl trimethylammonium to the negatively
charged silica.

\section{Introduction}

The study of the rheology of surfactants, polymers,
polyelectrolytes and their composites is interesting, not only
from an academic point of view, but also in the context of
possible industrial applications. Viscoelastic surfactant
solutions are ideal candidates for the study of complex flow
phenomena \cite{ch6rehage}, while mixtures of colloids and liquid
crystals form novel composite materials under certain conditions.
There has been extensive work on the dramatic changes in the
rheology of liquid crystals on the addition of colloidal particles
\cite{ch6zapotocky,ch6basappa,ch6meeker}. When latex microspheres
bind to flexible giant vesicles, the resulting membrane
distortions induce interparticle attractions and the subsequent
formation of close-packed and ring-like particle aggregates
\cite{ch6koltover}. Recent x-ray studies show that mixtures of DNA
and cationic or neutral liposomes form stacks of two-dimensional
smectics of aligned DNA chains intercalated by lipid bilayers
\cite{ch6radler}. The experiments of Solomon {\it et al.}
\cite{ch6solomon} prove that the yielding of concentrated zirconia
suspensions is extremely sensitive to the headgroup chemical
structure of the added surfactant. Phase separations in a mixture
of hard-sphere rods and colloids have been observed
experimentally, where spontaneous demixing into rod-rich and
rod-poor states, alternating layers of rods and spheres, columns
of spheres in a crystalline array etc. are seen to occur
\cite{ch6adams}. Recent experiments have shown a variety of
structures (2D hexagonal, lamellar and 2D centered rectangular
phase) formed by complexation of cationic and anionic  surfactant
mixtures with polyelecrolytes like DNA, poly-glutonic acid,
poly-acrylic acid and polystyrene sulfonate \cite{aksepl}.

In addition to the systems mentioned above, extensive experimental
investigations have been cconducted on mixtures of different
species of surfactants and polymers. Cetyl trimethylammonium
bromide (CTAB) micelles show a dramatic increase in length when
mixed with a few molecules of oppositely-charged gemini
surfactants \cite{ch6menger}. The results are explained in terms
of the cross-linking of the CTAB micelles by the gemini
surfactants. Viscometric measurements on mixtures of cetyl
trimethylammonium tosylate (CTAT) and cetyl trimethylammonium
3-hydroxy naphthalene 2- carboxylate (CTAHNC) show a peak in the
zero-shear viscosity $\eta_{\circ}$ with increasing HNC$^{-}$
concentration \cite{ch6hassan}. The added HNC$^{-}$ is thought to
increase the micellar hydrophobicity, which encourages micellar
growth and increases the solution viscosity. The subsequent
decrease in $\eta_{\circ}$ is understood in terms of the formation
of intermicellar connections. Addition of the anionic micelle
sodium dodecyl sulphate (SDS) to hydrophobically modified
poly(sodium acrylate) (HMPAA) also shows a peak in the viscosity
\cite{ch6iliopoulos}. A maximum in the viscosity of the polymer
poly(ethyleneoxide) (PEO) on the addition the surfactant SDS is
explained in terms of a polymer-surfactant complex formation
\cite{ch6chari}. This peak is shown to correspond to the
concentration of SDS at which the polymer coils get saturated with
the added surfactant. Addition of methyl-Na-benzoates to cetyl
pyridinium chloride (CPyCl) results in a peak in the zero-shear
viscosity \cite{ch6rehage}. This is explained in terms of a rapid,
micellar growth followed by cross-linking of the micelles in the
presence of the hydrophobic methyl-Na-benzoate. The stress
relaxation changes from multi-exponential to mono-exponential
across the viscosity peak, indicating a cross-over from
diffusion-controlled to kinetically-controlled processes. It is
thus easy to control the rheology of polymer and surfactant
solutions by changing the counterion concentration or by
introducing micron-sized colloidal additives. Addition of
hydrophobic counterions or colloidal particles modifies the
inter-micellar interactions and alters the micellar packing
considerations, and in some cases, may even result in phase
transitions and the formation of novel complexes.

In this paper, we discuss our work on the rheology of semi-dilute
solutions of wormlike micelles (cetyl trimethylammonium tosylate)
in the presence of sub-micrometer sized particulate additives
(monodisperse silica colloids of diameter 0.1 $\mu$m). In addition
to the frequency dependent elastic modulus G$^{\prime}(\omega)$
and the loss modulus G$^{\prime\prime}(\omega)$, we measure the
shear rate dependent viscosity $\eta(\dot\gamma)$ of the samples,
and by fitting these to appropriate models, extract the relaxation
times $\tau_{R}$, the zero shear viscosities $\eta_{\circ}$ and
the high frequency plateau moduli $G_{\circ}$ of the mixtures as a
function of the weight percent of added silica (SiO$_{2}$)
particles. We explain our observations by considering the changes
in inter-macromolecular interactions that arise out of the
presence of colloidal additives in the viscoelastic gel.

\section{Experimental}

CTAT powder, purchased from Sigma Chemicals, Bangalore, India, is
weighed carefully and dissolved at weight percents of 1.4wt.\%
(31mM), 1.95wt.\% (43mM) and 2.6wt.\% (53mM) in aqueous colloidal
silica suspensions prepared at weight percents 0.85wt.\%,
1.05wt.\%, 1.4wt.\%, 1.95wt.\%, 2.6wt.\%, 3.9wt.\% and 5.2wt.\%.
The silica colloids were obtained as a gift from Nissan Chemicals,
Japan. Pure samples of CTAT are also prepared at weight percents
1.4 wt.\%, 1.95 wt.\% and 2.6 wt.\% to understand the role of the
colloidal silica on the rheology of the viscoelastic gel phase of
CTAT. In this concentration range, the zero shear viscosity
$\eta_{\circ}$ of CTAT is found to vary with CTAT concentration
$c$ as $\eta_{\circ} \sim c^{3.8}$.

\noindent The CTAT - silica mixtures are allowed to equilibrate at
30$^{\circ}$C for a week. The silica suspensions in which CTAT is
dissolved are prepared from a stock solution of monodisperse
colloidal silica (diameter = 0.1 $\mu$m, density = 1.29 g/cc. and
weight fraction 39.3\%), diluted with deionized and distilled
water to the required concentrations. The effect of electrostatic
interactions on micellar aggregation in the presence of colloidal
SiO$_{2}$ is studied by increasing the surface charge on the
SiO$_{2}$ particles. This is achieved by adding minute quantities
of sodium hydroxide ([NaOH] $\sim$ 0.178mM) to the silica
suspensions. Addition of NaOH to SiO$_{2}$ results in the
increased dissociation of the surface silanol groups, which leads
to an increase in the surface charge density of the silica
particles \cite{ch6yamanaka}.

All the rheological measurements are performed in a cone-and-plate
geometry (cone angle = 1$^{\circ}$59$^{'}$, diameter = 4 cm) in an
AR-1000N Rheolyst stress-controlled rheometer (T. A. Instruments,
U.K). The sample temperature is controlled at 25$^{\circ}$C for
all the experiments. Direct visualization of the samples under
ambient conditions using optical microscopy shows no evidence of
phase separation.

\section{Linear rheology}

In viscoelastic gel-forming surfactant solutions, the stress
relaxation is governed by the breaking and reformation time
$\tau_{break}$ and the reptation or curvilinear diffusion time
$\tau_{rep}$ of the micelles. In the limit where $\tau_{break} <<
\tau_{rep}$, the resulting single exponential stress relaxation
can be fit to the Maxwell model, where the characteristic
relaxation time $\tau_{R} = \sqrt{\tau_{rep}\tau_{break}}$. Fig. 1
shows the frequency response data for CTAT 1.4wt.\% samples, with
(a) 0wt.\%, (b) 1.3wt.\% and (c) 5.2wt.\% silica. The elastic
modulus $G^{\prime}(\omega)$ and the viscous modulus
$G^{\prime\prime}(\omega)$ do not show pure Maxwellian behavior,
and are fit to the Cole-Davidson model \cite{ch6cole} which
predicts a stretched exponential form for the stress relaxation
and works well for many glass-forming liquids. In this model, the
dynamic modulus is expressed as $G^{\star}(\omega) = G_{\circ}[1 -
\frac{1}{(1+i\omega\tau_{R})^{\alpha}}]$, where $G_{\circ}$ is the
high frequency plateau modulus and $\tau_{R}$ is a measure of an
average relaxation time. The inverse of the exponent $\alpha$
characterizes the width of the relaxation spectrum and equals 1
for Maxwellian relaxation \cite{ch6cates1,ch6cates2}. A decrease
in $\alpha$ from its maximum value of 1 indicates the presence of
multiple stress relaxation processes. The frequency dependent
moduli measured in our experiments are fit to the Cole-Davidson
model with $G_{\circ}$, $\tau_{R}$ and $\alpha$ as the fitting
parameters. For angular frequencies $\omega <$ 2 rad/sec, the fits
(shown as solid lines in fig. 1) are found to improve as the
concentration of silica additives in the CTAT solutions is
increased. The deviations observed at high angular frequencies
($\omega >$ 2 rad/sec) can be attributed to the effects of
breathing (axial stretching modes arising from tube length
fluctuations) and Rouse modes \cite{granek,ch6kern1} on the stress
relaxation processes in the samples. Fig. 2 shows the plot of the
fitted values of $\alpha$ as a function of increasing SiO$_{2}$
content. The values of $\alpha$, also tabulated in table 1, are
found to increase monotonically with SiO$_{2}$ concentration,
implying a separation in the relaxation time scales and a tendency
towards increasingly mono-exponential relaxation on the addition
of SiO$_{2}$ particles. $\tau_{R}$ may also be calculated from the
crossover frequency $\omega_{co}$ at which G$^{\prime}(\omega)$
and $G^{\prime\prime}(\omega)$ are equal, by using the relation
$\tau_{R} = \omega_{co}^{-1}$. A good indication of single
exponential stress relaxation is the semi-circular form of the
so-called Cole-Cole figure, where the viscous modulus normalized
by the high frequency shear modulus $G_{\circ}$ is plotted {\it
vs.} the normalized elastic modulus. Fig. 3 shows the Cole-Cole
plot corresponding to the data plotted in Fig. 1. The plot for the
sample with 5.2 wt.\% SiO$_{2}$ in CTAT approaches the
semi-circular form indicative of the dominance of a unique stress
relaxation mechanism. The Cole-Cole plot for the pure 1.4 wt.\%
CTAT is almost linear and implies a distribution of relaxation
time scales, possibly arising out of competing processes (for
example, comparable breakage and reptation times). Figs. 4 (a),
(b) and (c) show the plots of the dynamic viscosity $\eta^{\star}$
normalized by the zero-shear viscosity $\eta_{\circ}$ (open
squares) for the same three samples, where $\eta^{\star}(\omega) =
\frac{\sqrt{G^{\prime 2}+G^{\prime\prime 2}}}{\omega}$. The dashed
lines show the fits to the model for giant wormlike micelles given
by $\eta^{\star}(\omega) =
\frac{\eta_{\circ}}{\sqrt{1+\omega^{2}\tau_{R}^{2}}}$
\cite{ch6fischer}, where $\tau_{R}$ is the average relaxation time
discussed earlier. The fits of $\eta^{\star}(\omega)$ to this
model are found to work best at intermediate SiO$_{2}$
concentrations. These results can be interpreted in terms of the
growth of entangled wormlike micelles, followed by the formation
of bilayers due to the adsorption of the headgroups on the
SiO$_{2}$ surfaces.

\section{Nonlinear rheology}

The zero-shear viscosities $\eta_{\circ}$ of all the samples are
also calculated by fitting the shear-viscosity
($\eta(\dot\gamma)$) {\it vs.} shear rate ($\dot\gamma$) data
obtained in the flow experiments to the Giesekus model
\cite{ch6giesekus}. The Giesekus model considers the orientation
effects of flow by introducing a deformation dependent tensorial
mobility of breaking and reforming micelles and predicts the
shear-rate dependent viscosity $\eta(\dot\gamma) =
\frac{\eta_{\circ}}{2\tau_{R}{\dot\gamma}^{2}}[\sqrt{1 +
4\tau_{R}{\dot\gamma^{2}}}-1]$. This expression is an analytical
solution of a constitutive equation for the viscoelastic
properties of entangled micelles and assumes an adjustable
parameter $\alpha^{\prime}$ = 0.5 \cite{ch6rehage}, where
$\alpha^{\prime}$ is a dimensionless anisotropy factor which is
related to the relative mobility tensor $\beta$ and the
configuration tensor $C$ by the relation $\beta = 1 +
{\alpha}(C-1)$. We would like to note here that experimental data
for wormlike micellar solutions usually satisfy $\alpha^{\prime}$
= 0.5 \cite{ch6rehage}. In addition to plots of
$\eta^{\star}(\omega)/\eta_{\circ}$ {\it vs.} $\omega$, fig. 4
also shows the plots of $\eta(\dot\gamma)/\eta_{\circ}$ (open
circles) {\it vs.} $\dot\gamma$ for the three samples and the fits
to the Giesekus model. A comparison of the values of
$\eta_{\circ}$ obtained from the fits to the dynamic viscosity and
shear viscosity data has been made in table 1. The fits to the
shear viscosity data are found to work best at intermediate
SiO$_{2}$ concentrations. Fig. 4 may be used to compare the degree
of validity of the Cox-Merz condition for these samples. The Cox
Merz rule is an empirical law for wormlike micelles which predicts
$\eta(\dot\gamma) = |\eta^{\star}(\omega)|$ at $\omega =
\dot\gamma$. We see that the agreement to the Cox-Merz rule is
very poor for the pure CTAT sample (fig. 4 (a)) and improves for
the intermediate (fig. 4 (b)) silica concentration, indicating the
growth of giant wormlike micelles. We also measure the first
normal force difference N$_{1}(\dot\gamma)$ generated in each
sample on the imposition of high shear rates and plot the results
in Fig. 5. The magnitude of the normal stress difference is found
to increase with SiO$_{2}$ concentration and indicates the
formation of structures on the addition of SiO$_{2}$.

All the experiments and analyses discussed above are repeated for
CTAT-SiO$_{2}$ mixtures with CTAT at concentrations of 1.4 wt.\%
(+ 0.178 mM of added NaOH added to increase the surface charge
density of SiO$_{2}$), 1.95 wt.\% and 2.6 wt.\% respectively. The
values of $\tau_{R}$, G$_{\circ}$ and $\eta_{\circ}$ calculated by
fitting the data to the models discussed above are plotted in
Figs. 6-10. In fig. 6, we plot the results for CTAT 1.4wt.\% +
SiO$_{2}$. In fig. 7, we compare the values of $\tau_{R}$,
G$_{\circ}$ and $\eta_{\circ}$ for CTAT 1.4wt.\% + SiO$_{2}$,
without added NaOH (solid circles) and with 0.178mM NaOH (open
triangles).  All the plots for the mixtures with untreated
SiO$_{2}$ show prominent peaks at SiO$_{2}$ weight percents
between 1.3 and 1.9. The peaks clearly shift to lower SiO$_{2}$
concentrations when SiO$_{2}$ is treated with NaOH. Figs. 8 and 9
are the plots of $\tau_{R}$, G$_{\circ}$ and $\eta_{\circ}$ for
CTAT-silica mixtures with the CTAT weight percent fixed at 1.95
and 2.6 respectively. In table 1, we have compiled the values of
the fitting parameters obtained from the fits to the frequency
response and steady shear experiments on CTAT 1.4 wt.\% as a
function of added SiO$_{2}$ concentration.
 \begin{table}
\begin{center}
\caption{The parameters $\alpha$, $G_{\circ}$, $\tau_{R}$ and
$\eta_{\circ}$, obtained by fitting the CTAT 1.4 wt.\% + SiO$_{2}$
data to the models described above. $\alpha$ and $G_{\circ}$ are
obtained from fits to the Cole Davidson model, the mean relaxation
time $\tau_{R}$ is the inverse of the crossover frequency
${\omega_{co}}^{-1}$ and $\eta_{\circ}$ is obtained by fitting to
the dynamic viscosity (DV) and the Giesekus (GM) models
respectively.} \vspace{0.5cm}
\begin{tabular}{|c|c|c||c|c|c|}\hline
{\rule[-3mm]{0mm}{8mm} SiO$_{2}$ wt.\%} &$\alpha$ &$G_{\circ}$
&$\tau_{R}$
 &$\eta_{\circ}$ Pa-s (DV) &$\eta_{\circ}$ Pa-s (GM)\\
\hline
\hline {\rule[-3mm]{0mm}{8mm} 0}& -  &1.3 &1.54 &2 &1\\
\hline {\rule[-3mm]{0mm}{8mm} 0.85}&$\sim$ 0.005  &1.3 &6.66 &8 &12\\
\hline {\rule[-3mm]{0mm}{8mm} 1.05}&$\sim$ 0.01  &1.5 &9.09 &12 &18\\
\hline {\rule[-3mm]{0mm}{8mm} 1.3}&0.01  &1.8 &11.11 &18 &21\\
\hline {\rule[-3mm]{0mm}{8mm} 1.95}&0.2 &2.1 &8.77 &19 &20\\
\hline {\rule[-3mm]{0mm}{8mm} 2.6}&0.3 &1.9 &5.40 &11 &12\\
\hline {\rule[-3mm]{0mm}{8mm} 3.9}& 0.6  &1.8 &3.57 &7 &8\\
\hline {\rule[-3mm]{0mm}{8mm} 5.2}& 0.7  &1.1 &2.00 &2 &2\\
\hline
\end{tabular}
\end{center}
\end{table}
The values of $\eta_{\circ}$ obtained from fits to the dynamic
viscosity and the Giesekus models are found to agree more at
intermediate SiO$_{2}$ concentrations, indicating a better
agreement with the empirical Cox-Merz rule under these conditions.
In fig. 10, we have plotted $\eta_{\circ}$,
 and $\tau_{R}$ for the different CTAT concentrations.
 For CTAT 1.4wt.\%, the values of $\eta_{\circ}$, $G_{\circ}$ (not shown) and
$\tau_{R}$ are found to peak at silica concentrations $\simeq$
CTAT concentrations. The peaks shift to lower SiO$_{2}$
concentrations for CTAT 1.95wt.\%. For CTAT 2.6wt.\%, the peaks
disappear or possibly shift to very low silica concentrations that
lie beyond the range of the present measurements.

\section{Discussions}
As discussed earlier, the stress relaxation in viscoelastic
solutions of wormlike micelles is dominated by two processes,
reptation and reversible scission
\cite{ch6rehage,ch6cates1,ch6cates2}. The value of $\alpha$ in the
Cole-Davidson model \cite{ch6cole} gives an estimate of the width
of the distribution of relaxation times. Multi-exponential
relaxation processes are seen in dilute micellar solutions where
$\alpha <$ 1, with $\alpha$ increasing to 1 with increasing
surfactant and/or salt concentrations \cite{ch6kern,ch6ranj}. On
adding SiO$_{2}$ to semi-dilute aqueous solutions of CTAT, we find
a similar increase in the values of $\alpha$ suggesting an
approach towards single exponential stress relaxation. From our
viscosity data, it is clear that there is an increased tendency
towards behavior characteristic of giant wormlike micelles,
followed by significant departures on further addition of
SiO$_{2}$ particles.

In this section, we explain the changes in the structure and
dynamics of the cylindrical CTAT micelles when silica colloids are
added to CTAT solutions under appropriate conditions. In the case
of CTAT 1.4 wt.\%, an addition of 1.3 wt.\% SiO$_{2}$ particles
results in the relaxation time increasing by 600\%, $G_{\circ}$
increasing by almost 37\% and $\eta_{\circ}$ increasing by 1600\%
as compared to pure CTAT solutions. We find that the peaks in
$\eta_{\circ}$ and $\tau_{R}$ shift to lower concentrations of
SiO$_{2}$ as the surfactant concentration is increased from 1.4 to
1.95 wt.\%, and disappear for CTAT 2.6wt.\% (fig. 10). The
increased fluidity of the samples on the addition of SiO$_{2}$ can
be explained by considering the adsorption of the surfactant
headgroups to the silica surfaces due to the attractive
interactions between them \cite{ch6favoriti}. The surfactants can
be thought to form bilayers around the silica surface, with the
headgroups constituting the outer layer and with the silica at the
center. Enhanced formation of bilayers with increasing SiO$_{2}$
concentration can therefore explain the observed decreases in
$\eta_{\circ}$, $G_{\circ}$ and $\tau_{R}$. At lower SiO$_{2}$
concentrations, giant wormlike micelles outnumber the bilayers
formed, and the presence of SiO$_{2}$ in solution serves to screen
the electrostatic interaction between micelles, resulting in
increased micellar entanglement, which manifests as peaks in plots
of the relaxation times, shear moduli and viscosities of the
samples. Addition of NaOH to 1.4wt.\% CTAT results in a decrease
in the magnitudes of $\eta_{\circ}$, $G_{\circ}$ and $\tau_{R}$,
with the peak shifting to lower SiO$_{2}$ concentrations (fig. 7).
Treating the SiO$_{2}$ particles with NaOH increases their surface
charge density by increasing the dissociation of the surface
silanol groups. Increasing the surface charge on silica causes
them to bind more strongly to the CTA$^{+}$, which results in the
formation of bilayers at lower silica concentrations, clearly
indicated by a shift in the peaks to the left (fig. 7). The
dramatic changes in the intermicellar interactions on treating the
SiO$_{2}$ particles with NaOH are thus responsible for the
observed changes in $\tau_{R}$, $G_{\circ}$ and $\eta_{\circ}$. On
continued addition of silica to the CTAT sample, a stage comes
when all the headgroups get adsorbed onto the silica surfaces,
leaving very few entangled micelles in solution.

Next, we focus on the shift in the viscosity peak to lower
concentrations of added silica, when the surfactant concentration
is increased to 1.95wt.\% (fig. 8) and 2.6 wt.\% (fig. 9). It is
useful to mention here that these weight percents lie in the
semi-dilute concentration regime of CTAT \cite{ch6soltero2}. An
increase in the surfactant concentration is known to result in an
increase in the degree of ionization of the micelles
\cite{ch6kern1}. Higher surfactant concentrations encourage the
formation of bilayers at lower silica concentrations, resulting in
the shifts of the peaks to lower silica concentrations. We
conclude this section by stressing that we identify the peak as a
signature of the formation of longer "worms" on the addition of
SiO$_{2}$. We believe that this is followed by the formation of
bilayers, resulting from the subsequent adsorption of the micellar
headgroups to silica, which is manifested experimentally as a
decrease in $\tau_{R}$, G$_{\circ}$ and $\eta_{\circ}$.

\section{Conclusions}
In this paper, we study the modifications in the rheology of
semi-dilute solutions of CTAT on the addition of submicrometer
sized silica spheres. The drastic changes in rheology that we
observe as a function of increasing silica concentration are
explained by considering the electrostatic interactions between
the surfactant and silica particles. We would like to note here
that modifications of the rheology of surfactant solutions by the
controlled addition of particulate matter can have diverse uses in
the industry.

\newpage

\begin{figure}
\caption{The elastic modulus $G^{\prime}(\omega)$ and the viscous
modulus $G^{\prime\prime}(\omega)$ {\it vs.} angular frequency
$\omega$ for CTAT 1.4wt.\%+SiO$_{2}$ samples. The solid lines show
the fits to the real and imaginary parts of the Cole-Davidson
model. The SiO$_{2}$ concentrations corresponding to each
frequency response curve are (a) 0 wt.\%, (b) 1.3wt.\% and (c)
5.2wt.\%.}
\end{figure}

\begin{figure}
\caption{Plot of the values of $\alpha$ {\it vs.} SiO$_{2}$
concentration for CTAT 1.4wt.\% solutions. $\alpha$ is obtained by
fitting the frequency response curves to the Cole-Davidson model.}
\end{figure}

\begin{figure}
\caption{Cole-Cole plots for CTAT 1.4 wt.\% with SiO$_{2}$
concentrations 0 wt.\% (open squares), 1.3 wt.\% (plus-centered
circles) and 5.2 wt.\% (filled diamonds) respectively.}
\end{figure}

\begin{figure}
\caption{The normalized dynamic viscosity
$\eta^{\star}(\omega)/\eta_{\circ}$ {\it vs.} $\omega$ (open
squares) and the normalized shear viscosity
$\eta(\dot\gamma)/\eta_{\circ}$ {\it vs.} shear rate (open
circles) ${\dot\gamma}$ for CTAT 1.4wt.\%+SiO$_{2}$ samples. The
dashed lines show the fits of the dynamic viscosity to
$\eta^{\star} = \frac{\eta_{\circ}}{\sqrt{1 +
\omega^{2}\tau_{R}^{2}}}$ and the solid lines show the fits to the
Giesekus model. The SiO$_{2}$ concentration corresponding to each
plot is (a) 0 wt.\%, (b) 1.3wt.\% and (c) 5.2wt.\%.}
\end{figure}

\begin{figure}
\caption{The first normal stress difference $N_{1}$ {\it vs.}
shear rate $\dot\gamma$ for CTAT 1.4wt.\%+SiO$_{2}$ samples.}
\end{figure}

\begin{figure}
\caption{Plot of viscoelastic parameters for CTAT 1.4wt.\%. (a)
shows the calculation of the average relaxation time $\tau_{R}$
from the crossover frequency (cross-centred triangles), fits to
the dynamic viscosity model (open circles) and the fits to the
Giesekus model (solid circles) respectively. (b) shows the plots
of the values of $G_{\circ}$ at different SiO$_{2}$ concentrations
obtained from the fits to the Cole-Davidson model. (c) shows the
values of $\eta_{\circ}$ obtained from the fits to the dynamic
viscosity model (open diamonds) and the Giesekus model (solid
circles).}
\end{figure}

\begin{figure}
\caption{Plots of the fitted parameters in the absence (filled
circles) and in the presence (open triangles) of 0.178mM NaOH for
CTAT 1.4wt.\%. (a) and (b) show the calculation of the zero-shear
viscosity $\eta_{\circ}$ from the fits to the Giesekus and the
dynamic viscosity models respectively and (c) shows the plots of
the values of $G_{\circ}$ at different SiO$_{2}$ concentrations
obtained from the Cole-Davidson fits. (d) shows the values of
$\tau_{R}$ obtained by using the relation $\tau_{R} \sim
\omega_{co}^{-1}$.}
\end{figure}

\begin{figure}
\caption{Plots of the fitted parameters for CTAT 1.95wt.\% +
SiO$_{2}$. (a) shows the plot of the average relaxation time
$\tau_{R}$ obtained from the inverse of the crossover frequency
(solid triangles), fits to the dynamic viscosity model (open
diamonds) and the fits to the Giesekus model (solid circles). (b)
shows the plots of the values of $G_{\circ}$ at different
SiO$_{2}$ concentrations obtained from the Cole-Davidson fits and
(c) shows the values of $\eta_{\circ}$ obtained from the fits to
the functional form suggested for $\eta^{\star}$ (open diamonds)
and the Giesekus model (filled circles).}
\end{figure}

\begin{figure}
\caption{Plots of the fitted parameters for CTAT 2.6wt.\% +
SiO$_{2}$. (a) shows the plot of the average relaxation time
$\tau_{R}$ i) obtained using the relation $\tau_{R} =
\omega_{co}^{-1}$ (solid triangles), ii) from fits to the dynamic
viscosity model (open diamonds) and (iii) from the fits to the
Giesekus model (solid circles), (b) shows a plot of the values of
$G_{\circ}$ at different SiO$_{2}$ concentrations, obtained from
the Cole-Davidson fits and (c) shows the values of $\eta_{\circ}$
obtained from the fits to the dynamic viscosity model (open
diamonds) and the Giesekus model (solid circles).}
\end{figure}

\begin{figure}
\caption{Plot of the zero-shear viscosity $\eta_{\circ}$ obtained
by fitting to the dynamic viscosity data for of 1.4wt.\%
(circles), 1.95wt.\% (up-triangles) and 2.6wt.\% (down-triangles)
CTAT. The relaxation times $\tau_{R}$ obtained from $\omega_{co}$
are also plotted {\it vs.} silica concentration for the three
concentrations of CTAT and are indicated by the same symbols
described above. The dotted lines are guides to the eye.}
\end{figure}
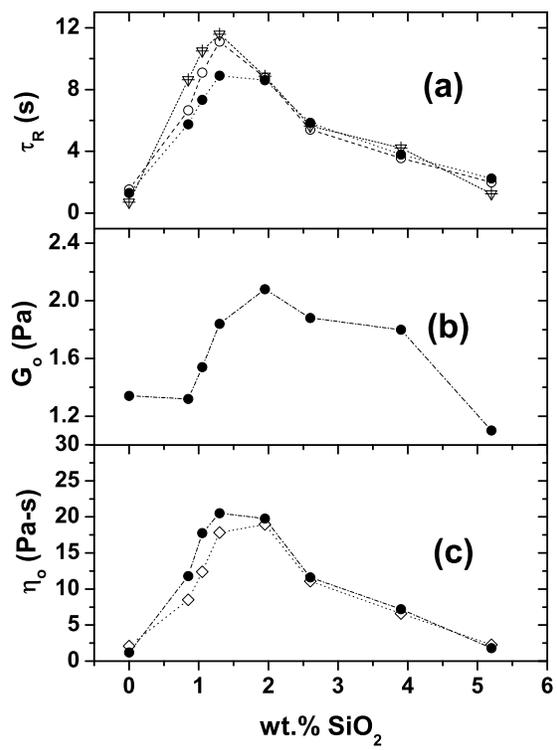
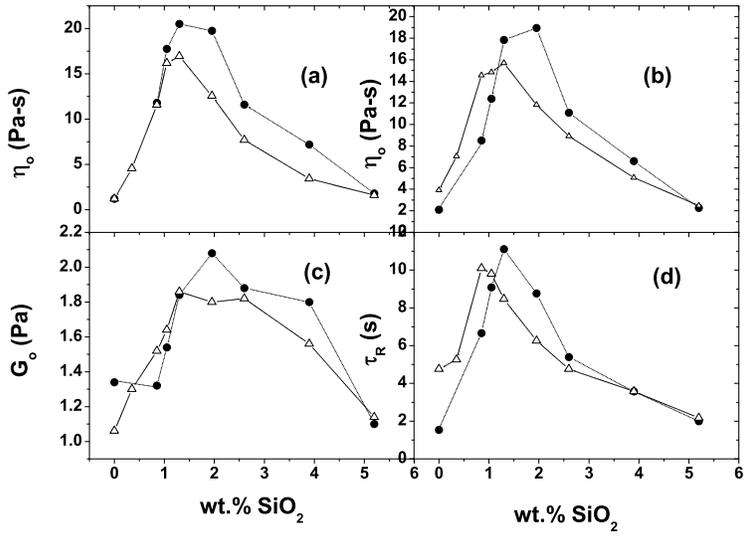
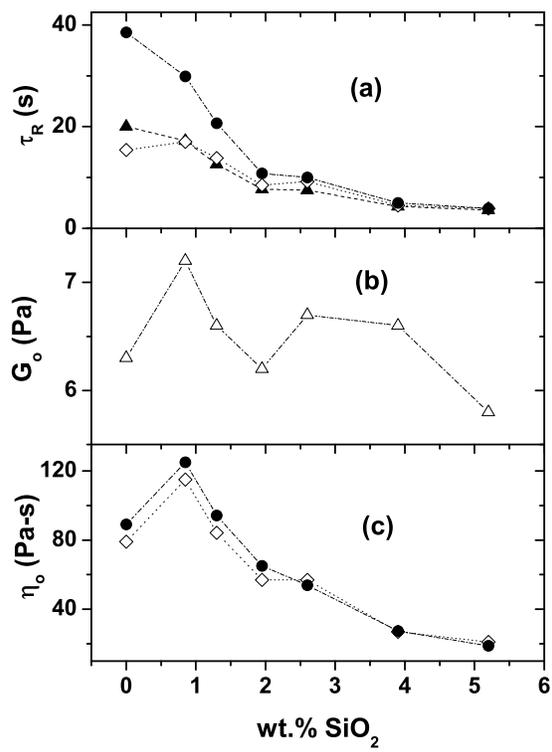
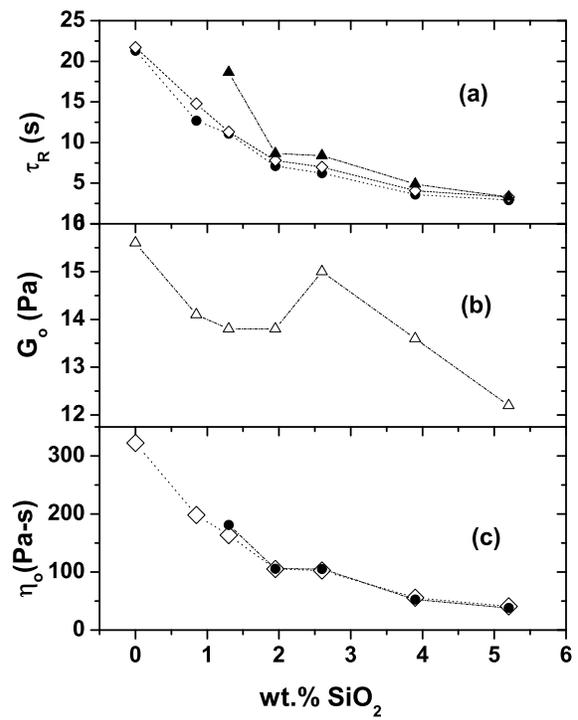
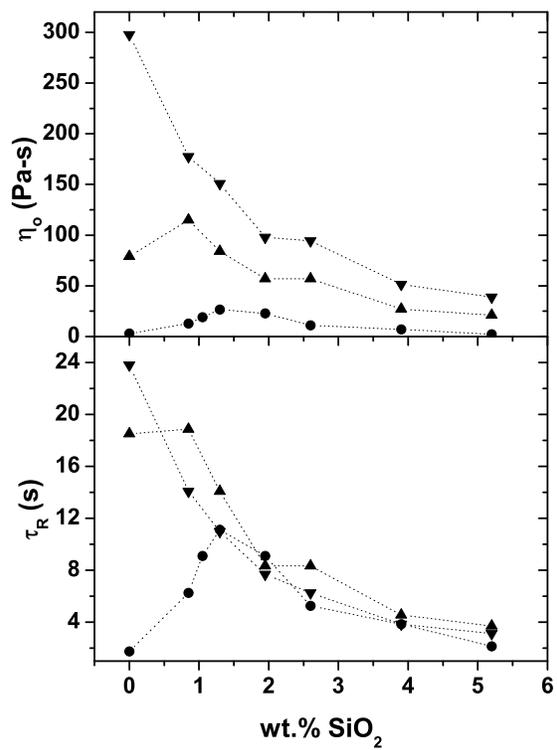
\end{document}